\documentclass[12pt]{iopart}
\usepackage{iopams}

\begin{document}
\title[Quasilocal energy in modified gravity]{Quasilocal 
energy in modified gravity}
\author{Valerio Faraoni $^1$}
\address{$^1$ Physics Department and {\em STAR} Research 
Cluster, Bishop's University, 2600 College Street, 
Sherbrooke, Qu\'ebec, Canada J1M~1Z7}
\ead{vfaraoni@ubishops.ca}
\begin{abstract}
A new generalization of the Hawking-Hayward quasilocal 
energy to scalar-tensor gravity is proposed without 
assuming symmetries, asymptotic flatness, or special 
spacetime metrics. The procedure followed is simple but 
powerful and consists of writing the scalar-tensor field 
equations as effective Einstein equations and then applying 
the standard definition of quasilocal mass. An alternative 
procedure using the Einstein frame representation leads to 
the same result {\em in vacuo}. 
\end{abstract}
\small
\begin{center} Keywords:~quasilocal energy, 
scalar-tensor gravity \end{center} \normalsize

\maketitle

\section{Introduction}
\label{sec:1}

There is little doubt that Einstein's General Relativity 
(GR) is not the final theory of gravity. GR breaks down at 
spacetime singularities and cannot be quantized. All the 
attempts to merge GR with quantum mechanics provide, in 
their low-energy limits, corrections to GR in the form of 
higher derivative equations or extra fields with explicit
 couplings to the spacetime curvature or to matter. A 
particularly compelling motivation for studying alternative 
theories of gravity comes from cosmology: the standard 
cosmological model based on GR, the $\Lambda$-Cold Dark 
Matter model, can only explain the present 
acceleration of 
the universe discovered with type Ia supernovae by invoking 
a completely {\em ad hoc} dark energy 
\cite{AmendolaTsujikawabook}.
 Perhaps we are already observing deviations from GR in the 
cosmic acceleration that the $\Lambda$-Cold Dark Matter 
model tries to 
fit into GR. Scalar-tensor theories of gravity and, in 
particular, the subclass known as $f({\cal R})$ gravity  
have enjoyed enormous popularity in the last decade 
\cite{Salvbook, reviews}, which only adds to previous 
motivation from string theory. In fact, 
the low-energy limit of string theories 
contains a dilaton very similar to the Brans-Dicke 
field of scalar-tensor gravity (the bosonic 
string theory reduces, in this limit, to a Brans-Dicke 
theory \cite{Fradkin}).  There is currently 
much interest in probing gravity at all scales  to detect 
or constrain deviations from GR (which could assume several 
forms in cosmology, black holes, or stellar 
objects \cite{Padilla, Bertietal2013}), including the 
search 
for 
scalar hair  \cite{scalarhair}.

The notion of mass of a relativistic gravitating system has 
been the subject of intense research in GR. Because of the 
equivalence principle, gravitational energy cannot be 
localized. The next best thing is 
a quasilocal notion of energy, {\em i.e.}, the energy 
contained in a compact 2-surface in spacetime, and several 
definitions of quasilocal energy have been introduced over 
the years (see \cite{Szabados} for a review). It seems that 
the relativity community settled on the Hawking-Hayward 
quasilocal construct \cite{Hawking, Hayward}, which we 
employ here but other quasilocal energies could be used as 
well.

The concept of mass is not only important in principle and 
for its obvious applications to gravitating systems, but 
also because it appears in the first law of thermodynamics 
for gravity. Much literature has been devoted to black hole 
thermodynamics and the thermodynamics of gravity and 
spacetime ({\em e.g.}, \cite{SptThermo}), but this is still 
an active area of theoretical research.

Given the significance of modified gravity 
\cite{Padilla, Bertietal2013, scalarhair} it would be 
important to know 
whether the 
quasilocal energy can somehow be extended to these 
theories, beginning with the simplest and most popular 
alternative, scalar-tensor gravity (see \cite{Hideki} for 
the case of $n$-dimensional Lovelock gravity). Thus 
far,  
discordant prescriptions for a quasilocal mass have been 
given \cite{Cai, Zhang, WuWangYang, Cognola} but they are 
subject to 
important restrictions: 1)~only $f({\cal R})$ gravity, 
which is a subclass of 
scalar-tensor theories, has been examined; 2)~only 
spherical symmetry, and sometimes only special 
spacetime geometries, have been considered. These 
prescriptions have been obtained using spacetime 
thermodynamics and a first law \cite{Cai, Zhang, 
WuWangYang, Cognola}. 
However, the expressions of the other four  
quantities used in the first law of 
thermodynamics (temperature, 
entropy, work density, and heat supply vector, 
respectively) are not established beyond doubt, which 
introduces some ambiguity in the definition of quasilocal 
mass obtained by {\em assuming} a certain form for the 
first law. 
Additionally, the concept of horizon 
temperature requires quantum considerations that are 
highly nontrivial in curved spacetime, where it is 
difficult to complete quantum field theory calculations 
unambigously. While we remain agnostic on these approaches 
in this paper, we propose to bypass these 
conceptual difficulties by 
introducing a quasilocal mass in scalar-tensor 
and $f({\cal R})$ gravity via considerations that are 
purely classical and independent of the thermodynamics of 
gravity. An advantage of this approach is that the 
generalization of the Hawking-Hayward mass to scalar-tensor 
gravity thus obtained is not restricted to the $f({\cal 
R})$ subclass nor to special metrics, and it does not 
require 
spherical symmetry or asymptotic flatness. To be 
concrete, we derive a quasilocal mass by  writing the 
scalar-tensor field equations as effective Einstein 
equations and using the geometric derivation of the 
Hawking-Hayward mass in this ``effective GR'' context.

Let us first review the basics of scalar-tensor gravity 
used here. The (Jordan frame) action is 
\begin{eqnarray}
S_{\mbox{ST}} = \int d^4 x \sqrt{-g} \left\{ \left[ 
\frac{1}{16\pi} \left(  \phi {\cal R} 
-\frac{\omega(\phi)}{\phi} \, g^{ab} \nabla_a \phi \nabla_b 
\phi \right) -V(\phi) \right]  
+{\cal L}_{(m)} \right\} \,,\nonumber\\
\label{Jframeaction} 
\end{eqnarray} 
where ${\cal R}$ is the Ricci curvature of 
the spacetime metric $g_{ab}$ with determinant $g$, $\phi$ 
is the Brans-Dicke-like scalar field (the inverse of the 
effective gravitational coupling strength 
$G_{\mbox{eff}}$, 
which is varying in these theories), $V(\phi)$ is a scalar 
field potential, and ${\cal L}_{(m)}$ is the matter 
Lagrangian density. The field equations of scalar-tensor 
theory are
\begin{eqnarray}
&& R_{ab}-\frac{1}{2} \, g_{ab}{\cal R} = 
\frac{8\pi}{\phi} \, T_{ab}  +\frac{\omega}{\phi^2} \left( \nabla_a\phi \nabla_b 
\phi - \frac{1}{2} \, g_{ab} \nabla^c \phi \nabla_c \phi 
\right) \nonumber\\ 
&\,& +\frac{1}{\phi} \left( \nabla_a\nabla_b \phi 
- g_{ab} \Box \phi \right) -\frac{V}{2\phi} \, g_{ab} 
\,,\label{STfieldeqs} \\
&&\nonumber\\
&&\Box \phi =  \frac{1}{2\omega+3} \left( 8\pi T 
-\frac{d\omega}{d\phi} \, \nabla^c\phi \nabla_c \phi +\phi 
\, \frac{dV}{d\phi} -2V \right)\,, \nonumber\\
&& 
\end{eqnarray}
where $T_{ab}=-\frac{2}{\sqrt{-g}} \frac{ \delta }{\delta 
g^{ab}} \left( \sqrt{-g}\, {\cal L}_{(m)}\right) $ is 
the 
stress-energy tensor of matter and $T \equiv {T^a}_a$. 
$f({\cal R})$ theories \cite{reviews} are a 
subclass of scalar-tensor theories of gravity described by 
the action
\begin{equation}
S = \int d^4 x \sqrt{-g} \, f({\cal R}) + 
S_{(m)}
\end{equation}
where $f({\cal R})$ is a nonlinear function of the Ricci 
scalar. By setting $\phi = f'({\cal R})$ and
\begin{equation}
V(\phi)= \phi {\cal R}(\phi) -f\left( {\cal R}(\phi)  
\right) \,,
\end{equation}
the action can be shown to be equivalent to the 
scalar-tensor one \cite{reviews}
\begin{equation}
S = \int d^4 x \, \frac{ \sqrt{-g}}{16\pi}  \left[ \phi 
{\cal R}-V(\phi) \right] +S_{(m)} \,,
\end{equation}
a Brans-Dicke action with vanishing Brans-Dicke parameter  
$\omega$ and potential $V$ for the Brans-Dicke scalar 
$\phi$.

\section{Scalar-tensor quasilocal mass}
\label{sec:2}

The Hawking-Hayward quasilocal mass is defined as follows 
\cite{Hawking, Hayward}: 
let ${\cal S}$ be an 
embedded spacelike, compact, and orientable 2-surface with 
induced 2-metric $h_{ab}$ and  induced Ricci scalar ${\cal 
R}^{(h)}$. Consider ingoing ($-$) and outgoing ($+$) null  
geodesic congruences from  ${\cal S}$. Let $\theta_{\pm}$ 
and $\sigma_{ab}^{\pm}$ be the expansions and shear 
tensors 
of these congruences, respectively, and $\omega^a$ be the 
projection onto ${\cal S}$ of the commutator of 
the null normal vectors to ${\cal S}$ (the 
anoholonomicity \cite{Hayward}). $\mu$ is the volume 
2-form on the surface ${\cal S}$ of area $A$. The 
Hawking-Hayward quasilocal energy is \cite{Hawking, 
Hayward}
\begin{equation}
M = \frac{1}{8\pi G} \sqrt{ \frac{A}{16\pi}} \int_{\cal S} \mu \left(
{\cal R}^{(h)} +\theta_{(+)} \theta_{(-)} -\frac{1}{2} \, \sigma_{ab}^{(+)} 
\sigma^{ab}_{(-)}   -2\omega_a\omega^a \right) \,. \label{HHmass}
\end{equation}
The contracted Gauss equation \cite{Hayward}
\begin{equation}
{\cal R}^{(h)} +\theta_{(+)} \theta_{(-)} -\frac{1}{2} \, \sigma_{ab}^{(+)} 
\sigma^{ab}_{(-)}  = h^{ac}h^{bd} R_{abcd}
\end{equation}
can be used to compute the first three terms in the integral. 
The usual splitting of the Riemann tensor into Weyl tensor and 
Ricci part
\begin{equation}
R_{abcd}=C_{abcd} + g_{a[c}R_{d]b} -g_{b[c} R_{d]a} -\frac{{\cal R}}{3} \, 
g_{a[c} g_{d]b} 
\end{equation}
and the effective Einstein equation~(\ref{STfieldeqs}) yield
\begin{eqnarray}
&&h^{ac} h^{bd} R_{abcd} = h^{ac} h^{bd} C_{abcd} \nonumber\\
&&\nonumber\\
& & +\frac{8\pi}{\phi} h^{ac} h^{bd} \left[ g_{a[c}T_{d]b} -g_{b[c}T_{d]a} 
-\frac{T}{2}  \left( g_{a[c}g_{d]b} -g_{b[c}g_{d]a} \right)\right] \nonumber\\
&&\nonumber\\
& & +\frac{\omega}{\phi^2} h^{ac} h^{bd} \left( 
g_{a[c}\nabla_{d]}\phi \nabla_b\phi 
-g_{b[c}\nabla_{d]}\phi \nabla_a\phi \right) \nonumber\\
&&\nonumber\\
& & +\frac{1}{\phi} h^{ac} h^{bd} \left(  
g_{a[c}\nabla_{d]} \nabla_b\phi 
-g_{b[c}\nabla_{d]} \nabla_a\phi \right) \nonumber\\
&&\nonumber\\
& & +\frac{ \left( \Box\phi +V \right)}{2\phi} h^{ac} 
h^{bd} 
\left( g_{a[c}g_{d]b} -g_{b[c}g_{d]a} \right) \nonumber\\
&&\nonumber\\
&& +\left( \frac{8\pi T}{3\phi} -\frac{\omega}{3\phi^2} \, 
\nabla^c\phi \nabla_c\phi  -\frac{\Box\phi}{\phi} -\frac{2V}{3\phi} 
\right) h^{ac} h^{bd}  g_{a[c}g_{d]b} \,. 
\nonumber\\
&& \label{questa}
\end{eqnarray}
By computing the individual terms
\begin{eqnarray}
&& h^{ac} h^{bd} \left( g_{a[c}T_{d]b} -g_{b[c}T_{d]a}  \right) 
= h^{ab}T_{ab} \,,\\
&&\nonumber\\
&& h^{ac} h^{bd} \left( 
g_{a[c}\nabla_{d]}\phi \nabla_b\phi 
-g_{b[c}\nabla_{d]}\phi \nabla_a\phi \right) =
h^{ab} \nabla_a\phi \nabla_b \phi \,,\nonumber\\
&&\\
&& h^{ac} h^{bd} \left(  
g_{a[c}\nabla_{d]} \nabla_b\phi 
-g_{b[c}\nabla_{d]} \nabla_a\phi \right)
= h^{ab}\nabla_a\nabla_b \phi \,,\\
&&\nonumber\\
&& h^{ac} h^{bd}\left( g_{a[c}g_{d]b} -g_{b[c}g_{d]a} \right) =2 \,,\\
&&\nonumber\\
&& h^{ac} h^{bd} g_{a[c}g_{d]b}  =1 \,,
\end{eqnarray}
and putting them together in eq.~(\ref{questa}), one 
obtains
\begin{eqnarray}
M_{\mbox{ST}} &=&  \frac{1}{8\pi} \sqrt{ \frac{A}{16\pi}} 
\int_{{\cal S}}\mu
\phi \left[  h^{ac} h^{bd} C_{abcd} -2\omega_a\omega^a 
+\frac{8\pi}{\phi} \, h^{ab}T_{ab}- \frac{16\pi T}{3\phi}  
\right.\nonumber\\
&&\nonumber\\
& \, & \left. +\frac{h^{ab}\nabla_a\nabla_b \phi}{\phi} 
+\frac{\omega}{\phi^2} \left( 
h^{ab}\nabla_a \phi \nabla_b \phi 
-\frac{1}{3}\, \nabla^c\phi \nabla_c \phi \right) 
+\frac{V}{3\phi} \right] \,,\label{generalresult}
\end{eqnarray}
where the $\phi$ factor in the first term on the right hand 
side is introduced by the replacement $G \rightarrow 
 G_{\mbox{eff}}$. Note that we moved $1/G$ inside the 
integral in eq.~(\ref{HHmass}) {\em before} replacing $G$ 
with $G_{\mbox{eff}}$, because otherwise a factor $\phi$ 
replacing $1/G$ would appear outside the integral in 
eq.~(\ref{generalresult}), making $M_{\mbox{ST}}$ a 
function 
on the surface ${\cal S}$ instead of a number specified 
once this surface is assigned. 

The factor $1/\phi=G_{\mbox{eff}}$ 
does not multiply all the terms in square brackets in the 
integrand 
of~(\ref{generalresult}) which 
compose the Hawking-Hayward mass in Einstein theory, but 
only the 
two terms 
containing $T_{ab}$ and its trace. Therefore, in general 
one does not expect to isolate the entire GR integrand 
divided by $\phi$ in the integral. However, 
eq.~(\ref{generalresult}) reduces to the standard GR 
expression \cite{Hawking, Hayward} in the GR limit in which 
$\phi$ becomes constant. If $T_{ab}$ describes a perfect 
fluid, $T_{ab}=\left( P+\rho \right) u_a u_b +P g_{ab}$, 
with the fluid 4-velocity $u^a$ normal to the 2-surface 
${\cal 
S}$ ({\em i.e.}, $h_{ab}u^b=0$), then it is
\begin{equation}
\frac{8\pi}{\phi} \, h^{ab}T_{ab}- \frac{16\pi T}{3\phi} = 
\frac{16\pi \rho}{3\phi} 
\end{equation}
and the quasilocal mass does not depend explicitly on the 
pressure (as remarked in \cite{Hayward, Haywardspherical} 
in the spherical case).

\section{Spherical symmetry}
\label{sec:3}

Let us specialize now to spherical symmetry, in which the 
Hawking-Hayward quasilocal mass construct reduces 
\cite{Haywardspherical} to the better known 
Misner-Sharp-Hernandez mass \cite{MSH}, which is defined by
\begin{equation} \label{MSH}
M_{\mbox{MSH}}=\frac{R}{2G} \left( 1- \nabla^c R \nabla_c R 
\right) \,,
\end{equation}
where $R$ is the areal radius. Let ${\cal S}$ be now a 
2-sphere of symmetry with induced metric $h_{ab}$ and write 
the line element as
\begin{equation}
ds^2=g_{00}dt^2 +g_{11} dR^2 +R^2 d\Omega_{(2)}^2 =
I_{ab}dx^a dx^b+h_{ab} dx^a dx^b 
\end{equation}
in coordinates $\left( t, R, \theta, \varphi \right)$,
where $I_{ab}=~$diag$\left( g_{00}, g_{11} \right)$,  
$h_{ab}=~$diag$\left( R^2, R^2 \sin^2 \theta \right)$, and 
$d\Omega_{(2)}^2=d\theta^2 + \sin^2\theta d\varphi^2 $ is 
the metric on the unit 2-sphere. Eq.~(\ref{generalresult})  
becomes
\begin{eqnarray}
M_{\mbox{ST}} &=&  \frac{\phi R^3}{4} 
\left[  h^{ac} h^{bd} C_{abcd} 
+\frac{8\pi}{\phi} \, h^{ab}T_{ab}- \frac{16\pi T}{3\phi} \right. \nonumber\\
&&\nonumber\\
&\, & \left. +\frac{\omega}{\phi^2} \left( h^{ab}\nabla_a \phi\nabla_b \phi 
-\frac{1}{3}\, \nabla^c\phi \nabla_c \phi \right) 
 +\frac{h^{ab}\nabla_a\nabla_b \phi}{\phi}  
+\frac{V}{3\phi} \right] \,.\label{sphericalresult}
\end{eqnarray}
Let us specialize now to cosmological metrics and then to 
the subclass of $f({\cal R})$ theories of gravity.

\subsection{Friedmann-Lema\^itre-Robertson-Walker geometry}

Consider as an application the spatially flat 
Friedmann-Lema\^itre-Robertson-Walker 
(FLRW) space sourced by a perfect fluid. 
The line element is
\begin{equation} 
ds^2=-dt^2 +a^2(t) \left( dr^2 +r^2 d\Omega_{(2)}^2 \right)
\end{equation}
and the areal radius is $R(t,r)=a(t)r$. One obtains easily
\begin{equation} 
M_{\mbox{ST}}= \frac{4\pi R^3}{3} \rho +
\frac{\phi R^3}{4} \left( \frac{\omega}{3} \, \frac{ 
\dot{\phi}^2}{\phi^2} 
-\frac{2H\dot{\phi}}{\phi} +\frac{V}{3\phi} \right) \,.
\end{equation}
By using the fact that $  \frac{R}{2} \left( 
1-g^{ab}\nabla_a R \nabla_b R \right) 
=H^2 R^3/2$, $h^{ab} \nabla_a \nabla_b \phi = -2H 
\dot{\phi}$, and the Hamiltonian constraint 
\begin{equation} \label{Hamconstraint}
H^2 = \frac{8\pi \rho}{3\phi} 
-H\, \frac{\dot{\phi}}{\phi} +\frac{\omega}{6} 
\left( \frac{ \dot{\phi}}{\phi} \right)^2 +\frac{V}{6\phi} 
\equiv \frac{8\pi \left( \rho +\rho_{\phi} \right)}{3\phi}
\,, 
\end{equation}
and replacing $G$ with $G_{\mbox{eff}}=\phi^{-1}$ leads to 
\begin{equation}
 M_{\mbox{ST}} = \frac{H^2R^3 \phi}{2}= 
\frac{4\pi R^3}{3} \left( \rho +\rho_{\phi} \right)
= \frac{R}{2} 
\left( 1-\nabla^c R\nabla_c R \right)\phi \,.
\end{equation}
This is nothing but the expression of the 
Misner-Sharp-Hernandez mass~(\ref{MSH}) with the 
replacement $G \rightarrow G_{\mbox{eff}}$. 

\subsection{FLRW space in $f({\cal R})$ gravity}

In metric $f({\cal R}) $ gravity it is $\phi=f'({\cal R})$, 
$\omega=0$, $V(\phi)= f'({\cal R}){\cal R}-f({\cal R})$ and,
using the analogue of eq.~(\ref{Hamconstraint}) 
\cite{reviews}
\begin{equation} 
H^2 = \frac{1}{3f'} \left[ 8\pi \rho +\frac{ 
{\cal R}f'-f}{2} 
-3H (f')\,\dot{} \right] \,,
\end{equation}
one obtains \cite{footnote}
\begin{equation} \label{sphericalgeneral}
M_{f({\cal R})}= \frac{H^2R^3\phi}{2}= 
\frac{4\pi R^3}{3} \, \rho
+\frac{R^3}{2} \left( \frac{ {\cal R} f'-f}{6} 
- H f'' \dot{{\cal R}}  \right) \,.
\end{equation}

\section{Einstein frame}
\label{sec:3bis}

It is possible to derive the result~(\ref{generalresult}) 
{\em in vacuo}  also with an independent procedure using 
the Einstein frame representation of scalar-tensor 
gravity, although the derivation relies heavily on a 
technical result about the transformation of the quasilocal 
mass under conformal rescalings which was obtaiend only 
recently \cite{AVVM}. 

As is well known, under the conformal rescaling of the 
metric 
\begin{equation}
g_{ab} \rightarrow \tilde{g}_{ab}=\Omega^2 g_{ab} \,, 
\;\;\;\;\;\; \Omega=\sqrt{\phi} \,,\label{delta1}
\end{equation}
and the non-linear scalar field redefinition $\phi 
\rightarrow \tilde{\phi}(\phi)$ with
\begin{equation}
d\tilde{\phi}=\sqrt{ \frac{2\omega +3}{16\pi}} \, 
\frac{d\phi}{\phi} \,, \label{delta2}
\end{equation}
the action~(\ref{Jframeaction}) assumes the Einstein frame 
form
\begin{equation}
S = \int d^4 x \sqrt{-\tilde{g}} \left[ \frac{ {\cal 
R}}{16\pi} 
-\frac{1}{2} \, \tilde{g}^{ab} \nabla_a \tilde{\phi} \, 
 \nabla_b \tilde{\phi}  -U ( \tilde{\phi})   
+\frac{ {\cal L}_{(m)}}{\phi^2} \right]  \,,
\end{equation}
where 
\begin{equation}
 U\left( \tilde{\phi}\right) =\frac{ V\left[ 
\phi(\tilde{\phi}) \right]}{  \left[  \phi ( 
\tilde{\phi}) \right]^2} \,.
\end{equation}
In the Einstein conformal frame with tilded variables 
$\left( \tilde{g}_{ab}, \tilde{\phi} \right)$ the ``new'' 
scalar field $\tilde{\phi}$ has canonical kinetic 
energy density and couples minimally with 
gravity but non-minimally with matter, hence the theory 
{\em in vacuo} is formally GR and the Hawking-Hayward 
quasilocal mass is well defined. The Einstein frame scalar 
field $\tilde{\phi}$ has canonical energy-momentum tensor
\begin{equation}
\tilde{T}_{ab}^{(\tilde{\phi})} = 
\nabla_a \tilde{\phi}\nabla_b \tilde{\phi} -\frac{1}{2} \, 
\tilde{g}_{ab}  \, \tilde{g}^{cd} \nabla_c \tilde{\phi}
\nabla_d \tilde{\phi} -U ( \tilde{\phi} ) 
\tilde{g}_{ab}  \,.
\end{equation}
Using eqs.~(\ref{delta1}) and (\ref{delta2}), this 
stress-energy tensor is written in terms of Jordan frame 
quantities as 
\begin{equation}
\tilde{T}_{ab}^{(\tilde{\phi})} = \frac{2\omega+3}{16\pi 
\phi^2} \left( \nabla_a \phi \nabla_b \phi -\frac{1}{2} \, 
g_{ab} \nabla^c \phi \nabla_c \phi \right) -\frac{V}{16\pi 
\phi} \, g_{ab} \label{delta4}
\end{equation}
and its Einstein frame trace is 
\begin{equation}
\tilde{g}^{ac} \tilde{T}_{ac}^{(\tilde{\phi})} = 
-\left( \frac{2\omega+3}{16\pi \phi^3}\right)   \nabla^c 
\phi 
\nabla_c \phi -\frac{V}{4\pi \phi^2} \,. \label{delta5}
\end{equation}
Regarding the scalar-tensor theory in the Einstein frame 
formally as GR (with the exception of the anomalous 
coupling 
of the scalar $\tilde{\phi}$ to matter which, as we shall 
see below, has some consequances), we can see matter in 
the Einstein frame as being described  by the total 
energy-momentum tensor
\begin{equation}
\tilde{T}_{ab}=\tilde{T}^{(m)}_{ab} 
+ \tilde{T}^{( \tilde{\phi})}_{ab} 
= \frac{T^{(m)}_{ab}}{\phi^2} +  
\tilde{T}^{( \tilde{\phi})}_{ab} \,. \label{delta5bis}
\end{equation}
The Hawking-Hayward quasilocal mass in the Einstein frame 
is then given by
\begin{eqnarray}
\tilde{M}_{ \mbox{GR}} &=& \frac{1}{8\pi} \sqrt{ \frac{ 
\tilde{A} }{16\pi} } \int_{ \tilde{ {\cal S} } } 
\tilde{\mu}\left[  \frac{1}{G} \left( \tilde{h}^{ac} 
\tilde{h}^{bd} 
\tilde{C}_{abcd}  +8\pi G \, \tilde{h}^{ab} 
\tilde{T}_{ab}^{(m)}  -\frac{16\pi G}{3}\, \tilde{g}^{ab} 
\tilde{T}_{ab}^{(m)} \right. \right.\nonumber\\
&&\nonumber\\
&\, & \left. \left.  +8\pi G \tilde{h}^{ab} 
\tilde{T}_{ab}^{(\tilde{\phi} )}  -\frac{ 16\pi G}{3} \, 
\tilde{g}^{ac} \tilde{T}_{ac}^{(\tilde{\phi}} 
\right) -2\tilde{\omega}_a\tilde{\omega}^a \right] \,, 
\label{delta6} 
\end{eqnarray}
being mindful of writing Newton's constant inside the 
integral in view of the discussion of the previous section. 
Since
\begin{eqnarray}
&& \tilde{h}^{ac} \tilde{h}^{bd} \tilde{C}_{abcd} = 
h^{ac} h^{bd} C_{abcd}/\phi \,, \label{delta8}\\
&&\nonumber\\
&& 8\pi G \tilde{h}^{ab} \tilde{T}_{ab}^{(m)}=
8\pi G h^{ab} \tilde{T}_{ab}^{(m)}/\phi \,,\label{delta9}\\
&&\nonumber\\
&&-\frac{16\pi G}{3}\, \tilde{g}^{ab} \tilde{T}_{ab}^{(m)}=
-\frac{16\pi G}{3\phi}\, \tilde{T}^{(m)} 
\,,\label{delta10}\\
&&\nonumber\\
&& 8\pi  \tilde{h}^{ab} \tilde{T}_{ab}^{( \tilde{\phi} )}=
\left( \frac{2\omega+3}{2\phi^3} \right) \, h^{ab} 
\nabla_a\phi \nabla_b \phi  -\left( 
\frac{2\omega+3}{2\phi^3} \right) \,
\nabla^c \phi \nabla_c \phi
 -\frac{V}{\phi^2} \,,\label{delta11}\\
&&\nonumber\\
&&-\frac{16\pi }{3}\, \tilde{g}^{ab} \tilde{T}_{ab}^{( 
\tilde{\phi} )} = \left( \frac{2\omega+3}{2\phi^3} \right) 
\, \nabla^c \phi \nabla_c \phi +\frac{4V}{3\phi^2}  \,, 
\label{delta12}
\end{eqnarray}
eq.~(\ref{delta6}) becomes
\begin{eqnarray}
\tilde{M}_{\mbox{GR}} &=& \frac{1}{8\pi} \sqrt{ \frac{ 
\tilde{A} }{16\pi} } \int\tilde{\mu}
\left[  \frac{ h^{ac} h^{bd} }{G}\, 
C_{abcd}  +8\pi h^{ab} \tilde{T}_{ab}^{(m)} 
-\frac{16\pi}{3}\,  
\tilde{T}^{(m)} \right.\nonumber\\
&&\nonumber\\
&\, & \left.  + \left( \frac{2\omega+3}{2\phi^2} \right) 
h^{ab}\nabla_a\phi \nabla_b \phi  -\left( 
\frac{2\omega+3}{6\phi^2} \right)
\nabla^c \phi \nabla_c \phi +\frac{V}{3\phi} 
-2\tilde{\omega}_a\tilde{\omega}^a \right] \,. \nonumber\\
&& \label{delta13} 
\end{eqnarray}
We can now {\em impose} that, under a conformal 
transformation from the Jordan frame to the Einstein frame, 
the quasilocal mass transforms as it does under conformal 
rescalings in GR and we {\em define} the quasilocal mass in 
the Jordan frame of scalar-tensor gravity according to this 
rule. This approach is independent of that of the 
previous section. The transformation rule of the quasilocal 
mass under conformal rescalings $g_{ab} \rightarrow 
\tilde{g}_{ab}=\Omega^2 g_{ab}$ in GR was obtained 
recently in \cite{AVVM} and is
\begin{eqnarray}
\tilde{M}_{ \mbox{HH} } &=&\sqrt{ \frac{\tilde{A} }{A}} \, 
M_{\mbox{HH}} \nonumber\\
&&\nonumber\\
&\, & +\frac{1}{4\pi} \sqrt{ \frac{ 
\tilde{A}}{16\pi}} \int_{ {\cal S} } \mu \left[ h^{ab} 
\left( \frac{2\nabla_a \Omega  \nabla_b \Omega}{\Omega^2} - 
\frac{\nabla_a \nabla_b \Omega}{\Omega} \right) -
\frac{\nabla^c \Omega  \nabla_c \Omega}{\Omega^2} 
\right]\,.\nonumber\\
&& \label{delta14}
\end{eqnarray}
In the special case of spherical symmetry, this formula 
reduces to the transformation property of the 
Misner-Sharp-Hernandez mass reporteds in \cite{VV}. Here we 
identify $\tilde{M}_{\mbox{HH}}$ with the quantity 
$\tilde{M}_{\mbox{GR}}$ of eq.~(\ref{delta13}) and 
$M_{\mbox{HH}}$ with the sought-for quasilocal mass in 
Jordan frame scalar-tensor gravity $M_{\mbox{ST}}$. Then 
one has
\begin{eqnarray}
M_{\mbox{ST}} &=& \sqrt{ \frac{A}{ \tilde{A} } } \, 
\tilde{M}_{\mbox{GR}} -\frac{1}{4\pi} \sqrt{ 
\frac{A}{16\pi}} 
\int_{ {\cal S}} \mu \left[ h^{ab}\left( 
\frac{2\nabla_a \Omega  \nabla_b \Omega}{\Omega^2} 
- \frac{\nabla_a \nabla_b \Omega}{\Omega} \right) 
\right.\nonumber\\
&&\nonumber\\
&\, & \left. - \frac{ g^{ab} \nabla_a\phi  \nabla_b \phi 
}{4\phi^2}  \right] \label{delta15} 
\end{eqnarray}
or, using $\Omega=\sqrt{\phi}$,
\begin{eqnarray}
M_{\mbox{ST}} &=& \sqrt{ \frac{A}{ \tilde{A} } } \,
\tilde{M}_{\mbox{GR}} -\frac{1}{4\pi} \sqrt{ 
\frac{A}{16\pi}}
\int_{ {\cal S}} \mu \left[ \frac{h^{ab}}{2\phi} \left(
\frac{3\nabla_a \phi  \nabla_b \phi}{2\phi} 
 -\nabla_a \nabla_b \phi  \right) \right.\nonumber\\
&&\nonumber\\
&\, & \left. - \frac{\nabla^c \Omega  \nabla_c 
\Omega}{\Omega^2} \right] \,. \label{delta16}
\end{eqnarray}
Eq.~(\ref{delta13}) then gives \cite{footnote2} 
\begin{eqnarray}
M_{\mbox{ST}} &=& \frac{1}{8\pi} \sqrt{ \frac{A}{16\pi}} 
\int \mu \left[ \frac{h^{ac}h^{bd} 
C_{abcd}}{G}-2\omega_a\omega^a +8\pi h^{ab} 
\tilde{T}_{ab}^{(m)} -\frac{16\pi \tilde{T}^{(m)}}{3} 
\right.\nonumber\\
&&\nonumber\\
&\, & \left.  + \frac{\left( 2\omega+3 \right)}{2\phi^2} \, 
h^{ab} \nabla_a 
\phi \nabla_b \phi +\frac{V}{3\phi} -\frac{ \left( 
2\omega+3\right)}{6\phi^2} \, \nabla^c\phi \nabla_c \phi 
\right.\nonumber\\
&&\nonumber\\
&\, & \left. 
-\frac{3}{2\phi^2} h^{ab} \nabla_a\phi \nabla_b \phi 
+ \frac{ h^{ab} \nabla_a\nabla_b \phi}{\phi} +\frac{ 
\nabla^c\phi \nabla_c\phi}{2\phi^2} \right] \,.
\end{eqnarray}
In the Jordan frame one replaces $G$ with 
$G_{\mbox{eff}}=\phi^{-1}$, which yields
\begin{eqnarray}
M_{\mbox{ST}} &=& \frac{1}{8\pi} \sqrt{ \frac{A}{16\pi}} 
\int \mu \left[ \phi h^{ac}h^{bd} 
C_{abcd} -2\omega_a\omega^a +8\pi h^{ab} 
\tilde{T}_{ab}^{(m)} -\frac{16\pi \tilde{T}^{(m)}}{3} 
\right.\nonumber\\
&&\nonumber\\
&\, & \left.  + \frac{\omega}{\phi^2} \, 
h^{ab} \nabla_a \phi \nabla_b \phi 
+\frac{ h^{ab} \nabla_a\nabla_b \phi}{\phi^2} 
-\frac{ \omega \nabla^c\phi \nabla_c \phi}{3\phi^2}   
+\frac{V}{3\phi} \right] \,.\label{muu}
\end{eqnarray}
{\em In vacuo}, this equation coincides with the 
result~(\ref{generalresult}) of the previous section but, 
in the presence of matter, 
$\tilde{T}_{ab}^{(m)}=T_{ab}^{(m)}/\phi^2 $ appears instead 
of $T_{ab}^{(m)}$. A possible explanation for this 
incomplete match between the two results~(\ref{muu}) 
and~(\ref{generalresult}) is that, formally,  
Einstein frame scalar-tensor gravity is not exactly GR 
because 
of the nonminimal coupling of the 
transformed Brans-Dicke-like scalar $\tilde{\phi}$ 
to matter, and therefore the definition of Hawking-Hayward 
mass is not completely appropriate, which leaves a memory 
in the translation of the mass $\tilde{M}_{\mbox{GR}}$ to 
the 
Jordan frame. In other words, the Einstein frame method is 
not fully applicable. However, in the absence of matter,  
Einstein frame scalar-tensor theory is formally GR 
with an ordinary scalar field minimally coupled and with 
canonical kinetic energy density, and the method does work. 
It should  be added that, in any case, the Einstein frame 
method requires the additional 
result~(\ref{delta14}) of Ref.~\cite{AVVM}, the derivation 
of which is highly non-trivial.

\section{Conclusions}
\label{sec:4}

We have derived a new formula for a quasilocal mass in 
general scalar-tensor gravity without assuming symmetries 
or asymptotic flatness and without restricting to a 
subclass of theories or to specific spacetime metrics. The 
avenue followed is simply to rewrite the scalar-tensor 
field equations as effective Einstein equations by 
regarding the $\phi$-dependent terms as an extra effective 
stress-energy tensor in their right hand side and by 
replacing Newton's constant $G$ with the varying coupling 
$G_{\mbox{eff}}=1/\phi$, as familiar in scalar-tensor 
gravity. This straightforward but powerful approach (which 
has been used successfully, for example, in cosmological 
perturbation theory \cite{Hwang} or in the initial value 
problem \cite{Salgado}) is completely independent of 
thermodynamical considerations and has the advantage that 
one does not need to guess, or derive expressions for, the 
thermodynamical quantities appearing in the first law, 
which are subject to a certain degree of ambiguity (see, 
{\em e.g.}, the discussion in \cite{mylastbook}). 
A second approach using the Einstein frame representation 
of scalar-tensor gravity,  and relying on a previous result 
on the conformal  transformation of the Hawking-Hayward 
mass in GR, reproduces the same result {\em in vacuo} but 
there is a difference in the presence of matter, probably 
because in this case Einstein frame scalar-tensor gravity 
is not formally GR.

Here we remain agnostic on the thermodynamic approach to 
the 
quasilocal mass in scalar-tensor gravity. However, when 
specialized to spherical symmetry and to FLRW space, our 
quasilocal mass proposal differs from previous 
prescriptions derived from a first law of thermodynamics 
for gravity (which also differ between themselves). This 
disagreement provides an independent approach to revisit 
the first law, which will be the subject of future work.

\ack 
I am grateful to Fay\c{c}al 
Hammad for pointing out typographical errors in a previous 
version of the manuscript and to Bishop's University 
and the Natural Sciences and Engineering Research Council 
of Canada for financial support. This research was 
supported by Perimeter Institute for Theoretical Physics. 
Research at Perimeter Institute is supported by the 
Government of Canada through Industry Canada
and by the Province of Ontario through the Ministry of 
Economic Development and Innovation.


\section*{References}

\begin{thebibliography}{999}

\bibitem{AmendolaTsujikawabook} Amendola L and  
Tsujikawa S 2010 {\em Dark Energy, Theory and 
Observations} (Cambridge: Cambridge Univ. Press)

\bibitem{Salvbook} Capozziello S and  Faraoni V 2010 {\em 
Beyond Einstein Gravity} (New York: Springer)

\bibitem{reviews} Sotiriou T P and Faraoni V 2010 {\em 
Rev. Mod. Phys.} {\bf 82} 451; De Felice A and Tsujikawa S 
2010, {\em Living Rev. Relativity} {\bf 13} 3;
Nojiri S and Odintsov S D 2011, {\em Phys. Rept.} 
{\bf 505} 59 

\bibitem{Fradkin} Callan C G,  Friedan D,  
Martinec, E J and Perry M J 1985 {\em Nucl. Phys.
B} {\bf 262} 593; Fradkin E S and Tseytlin A A 1985, 
{\em Nucl. Phys. B} {\bf 261} 1 

\bibitem{Padilla} 
Psaltis D, Perrodin D, Dienes K R and  
Mocioiu I 2008 {\em Phys. Rev. Lett.} {\bf 100} 
091101; {\em Erratum} 2008, {\bf  100}  119902; 
Jain B and Khoury J 2010 {\em Ann. Phys.
(NY)} {\bf 325} 1479; 
Clifton T, Ferreira P G, Padilla A  
and Skordis C 2012 {\em Phys. Rep.} {\bf 513} 1; 
Berti E, Cardoso V, Gualtieri L,  
Horbatsch M and Sperhake U 2013 {\em Phys. Rev. D} {\bf 
87} 124020; 
Baker T, Psaltis D and Skordis C 2015 {\em Astrophys. 
J.} {\bf 802} 63 

\bibitem{Bertietal2013} Berti E {\em et al.}, 
arXiv:1501.07274.

\bibitem{scalarhair} 
Jacobson T 1999 {\em Phys. Rev. Lett.} {\bf 83}, 2699;
Sotiriou T P and Faraoni V 2010 {\em Phys. Rev. Lett.} 
{\bf 108} 081103;
Horbatsch M W and Burgess C P 2012 {\em J. Cosmol. 
Astropart. Phys.} {\bf 1205} 010;
Cardoso V, Carucci I P,  Pani P and  
Sotiriou T P 2013 {\em Phys. Rev. Lett.} {\bf 111} 111101;
Herdeiro C A R and Radu E 2014 {\em Phys. Rev. Lett.} 
{\bf 112} 221101;
Sotiriou T P and Zhou S-Y  2014 {\em Phys. Rev. Lett.} 
{\bf  112} 251102

\bibitem{Szabados} Szabados L B 2009 {\em Living Rev. 
Relativity} {\bf 12} 4 

\bibitem{Hawking} Hawking S 1968, {\em J. Math. Phys. 
(N.Y.)} {\bf 9} 598 

\bibitem{Hayward} Hayward S A 1994 {\em Phys. Rev. D} 
{\bf 49} 831 

\bibitem{SptThermo} 
Jacobson T 1995 {\em Phys. Rev. Lett.} {\bf 75} 1260;
Eling C, Guedens R and Jacobson T 2006 {\em Phys. 
Rev. Lett.} {\bf 96} 121301; 
Hayward S A, Mukohyama S and Ashworth M C 1999,
{\em Phys. Lett. A} {\bf 256} 347;
Mukohyama S and Hayward S A 2000 {\em Class. 
Quantum Grav.} {\bf 17} 2153;
Cai R G and Kim S P 2005 {\em J. High Energy Phys.} 
{\bf  02} 050;
Cai R G and Cao L M 2007 {\em Phys. Rev. D} {\bf 75} 
064008;
Sheykhi A, Wang B and  Cai R G 2007 {\em Phys. Rev. D} 
{\bf  76} 023515; 
Cai R G, Cao L M and Hu Y P 2008 {\em J. High Energy 
Phys.} {\bf 08} 090;
Gong Y and Wang A 2007 {\em Phys. Rev. Lett.} {\bf 99}  
211301; 
Wu S F, Wang B, Yang G H  and Zhang P M 2008 
{\em Class. Quantum Grav.} {\bf 25} 235018;
Bamba K and Geng C Q 2009 {\em Phys. Lett. B} {\bf  679} 
282;
Akbar M and Cai R G 2007 {\em Phys. Rev. D} {\bf  75} 
084003;
Padmanabhan T 2002 {\em Class. Quantum Grav.} {\bf 19} 
5387; 2005 {\em Phys. Rep.} {\bf 406} 49; 
Paranjape A, Sarkar S and Padmanabhan T 2006 {\em Phys. 
Rev. D} {\bf 74} 104015; 
Kothawala D, Sarkar S and Padmanabhan T 2007 {\em Phys. 
Lett. B} {\bf 652} 338;
Chirco G, Haggard H M, Riello A and Rovelli C 2014 
{\em Phys. Rev. D} {\bf 90} 044044 

\bibitem{Hideki} Maeda H 2006{\em Phys. Rev. D} {\bf 
73} 104004; Maeda H and Nozawa M 2008 {\em Phys. Rev. D} 
{\bf 77} 064031 

\bibitem{Cai} Cai R G, Cao L M, Hu Y P and Ohta N 2009  
{\em Phys. Rev. D} {\bf 80} 104016; 
Cai R G, Cao L M, Hu Y P and Kim S P 2008 {\em Phys. Rev. 
D} {\bf 78} 124012 

\bibitem{Zhang} Zhang H, Hu Y and Li X-Z 2014 {\em 
Phys. Rev. D} {\bf 90} 024062 

\bibitem{WuWangYang} Wu S-F, Wang B and Yang G-H 2008, 
{\em Nucl. Phys. B} {\bf 799} 330  

\bibitem{Cognola} Cognola G, Gorbunova O, Sebastiani L 
and Zerbini S 2011  {\em Phys. Rev. D} {\bf 84} 023515 

\bibitem{Haywardspherical} Hayward S A 1996 {\em Phys. 
Rev. D} {\bf 53} 1938 

\bibitem{MSH} Misner C W and Sharp D H 1964 {\em Phys. 
Rev.} {\bf 136} B571; Hernandez W C and Misner C W 1966 
{\em Astrophys. J.} {\bf 143} 452
 
\bibitem{AVVM} Prain A,  Vitagliano, V,  Faraoni, V and 
Lapierre-L\'eonard, M  arXiv:1501.02977

\bibitem{VV} Faraoni V and Vitagliano V 2014 {\em Phys. 
Rev. D} {\bf 89} 064015 

\bibitem{footnote} This expression disagrees with that of 
\cite{Zhang} which contains extra terms and in which the 
sign of the last two terms on the right hand side of our 
eq.~(\ref{sphericalgeneral}) is the opposite of ours, which 
means that in \cite{Zhang} the density $\rho_{\phi}$ 
defined in eq.~(\ref{Hamconstraint}) is subtracted, instead 
of being added, to $ M_{f({\cal R})}$.

\bibitem{footnote2} There is no 
prescription for the conformal transformation of the 
quantity $\omega_a \omega^a $, which needs to be redefined 
in each conformal frame according to the normalization 
chosen for the 
4-tangents to the null geodesics congruences---see the 
discussion in Ref.~\cite{AVVM}.

\bibitem{Hwang} Hwang J C 1990 {\em Phys. Rev. D} 
{\bf 42} 2601; 1990 {\em Class. Quantum Grav.} {\bf 7} 
1613; 
1991 {\bf 8} 195; 1997 {\bf 14} 3327; 
Hwang J C and Noh H 1996 {\em Phys. Rev. D} {\bf 54} 1460

\bibitem{Salgado} Salgado M 2006 {\em Class. Quantum 
Grav.} {\bf 23} 4719 

\bibitem{mylastbook} Faraoni  V 2015 {\em Cosmological and 
Black Hole Apparent Horizons} (New York: Springer)


\end{thebibliography}
\end{document}